# Inverse problem for Planck formula

A. N. Pechenkov

pechenkov@imp.uran.ru

Planck formula is considered as a first moment (average value) of unknown function of electromagnetic energy distribution of black body radiation. Inverse problem for the definition of the unknown function is solved for Gibbs ensemble. The solution needs of ensembles with both absolute temperatures: positive temperature and negative temperature. Such ensembles are the part of more extended class of ensembles with finite energies and finite phase volumes. In addition, the absence of Bohr – van Leeuwen paradox is considered for such statistical ensembles.

**1. Canonical Gibbs ensemble with finite energy.**

Let me, please, remind briefly the Gibbs ensemble definition, follow by [1]. Let there is a number of physical systems ('elements' of the ensemble). Every system is interacting only with a thermostat, and not with other system ('independent' systems). The full energy of every system is random value.

There are three conditions for such thermal equilibrium ensemble. First condition is: full energy of the ensemble is the sum of the energies of all systems:

$$E = \sum_{n=1}^{N} E_n \qquad (1)$$

Second condition for independent random values is: the energy distribution of the ensemble is the product of the energy distributions of all systems:

$$dw = dw_1 dw_1 ... dw_N \qquad (2)$$

Third condition is: the energy distribution of the ensemble depends only from the ensemble energy, and not from the system energies.

All three conditions will be satisfied, if energy distributions will have follow view:

$$dw_n = a_n e^{bE_n} dE_n; \qquad dw = A e^{bE} dE; \quad A = \prod_n a_n; \; dE = \prod_n dE_n \qquad (3)$$

The energy distribution must be normalized:

$$\int_E dw = 1; \qquad \Rightarrow \quad A = \frac{1}{\int_E e^{bE} dE} \qquad (4)$$

The equation (4) must be discussed in more details [1, 2]. Usually we take follow assumption: $b = -1/kT < 0$. Here $k$ is Boltzmann coefficient, and $T$ is absolute temperature. We need of the assumption for the convergence of the integral in the equation (4), in the case of infinite large possible energy of the ensemble. According to (1), infinite large energy can be reached by two ways. First way consists in a possibility of infinite large energy of some elements of an ensemble. Second way consists in a possibility of infinite large number of elements of an ensemble.

We believe, full energy of an element can be infinite large, because of infinite large kinetic energy, which is a part of full energy.

But, the Nature hasn't any physical system with infinite large energy (except of electric self energy of point charges). Consequently, the ensembles with finite energy are more realistic ensembles, than the ensembles with infinite large possible energy. For such ensembles the convergence of the integral in the equation (4) will take place also for positive coefficient $b = 1/kT > 0$. It is unusual case, because large energies are more probable, than small energies of elements.

The sign of the coefficient $b$ must be determined by external physical conditions for Gibbs ensemble. For example, as it will be shown below, the sign will be determined by inverse problem solution for black body radiation.

Note here, that thermal equilibrium ensembles with more probable values of large energies are known in contemporary statistical physics [3 – 5]. They named as ensembles at 'negative absolute temperature'. The term is '…unfortunate and misleading…' [5], because we have only positive sign of coefficient $b$, but not negative sign of $T$. There is very small number of such ensembles. They are ensembles without of kinetic energy. For example, they are ensembles of nuclear magnetic spin systems in some crystals.

Emphasize again, in the article author believe, that finite energy and finite phase volume are more realistic properties of physical ensembles. Mathematically it means possibility of both: 'usual' and 'unusual' energy distributions for every ensemble. For every 'mathematical' distribution we must find a corresponding physical ensemble. The energy distribution must be determined by additional physical conditions.

1. **Inverse problem for Planck formula.**

As it is well known, Planck could not derive he's formula in the frame of classical continuous energy distribution of electromagnetic waves in black body radiation. Impossibility of such derivation is the postulate in all quantum mechanics literature. But, the postulate contradicts to inverse problem theory for moments of probability distributions.

Planck formula is first moment (average value) of wave energy distribution. Therefore, there is infinite number of energy distributions, which have the formula as first moment. The choice of only distribution must be governed by additional physical conditions. The choice can be examined by comparison of high order theoretical moments with corresponding experimental moments.

Planck formula for average energy of electromagnetic waves of black body radiation in a thermostat volume $V$ is [6]:

$$dE = \frac{8\pi v^2 V}{c^3} \frac{hv}{e^{\frac{hv}{kT}} - 1} dv \qquad (5)$$

Here: $T$ is absolute temperature of a thermostat; $v$ is wave frequency; $h$ is Planck constant; $c$ is vacuum light velocity.

Term $\frac{8\pi v^2 V}{c^3} dv$ is a number of standing waves in a volume $V$, in a band frequency $dv$, near a frequency $v$. Second term in equation (5) is the average energy of wave at a frequency $v$:

$$\overline{\varepsilon} = \frac{hv}{e^{\frac{hv}{kT}} - 1} \qquad (6)$$

Let all electromagnetic waves are independent one from another. Then we will use Gibbs ensembles. Let there is an upper limit of wave energy at every frequency, i.e. $\varepsilon(\nu,T) \leq \varepsilon_m(\nu,T)$. Gibbs energy distribution of waves at frequency $\nu$ is:

$$w(x,a,b) = ae^{bx}; x = \frac{\varepsilon}{\varepsilon_m} \in [0,1] \tag{7}$$

Inverse problem consist in determination of unknown coefficients (*a, b*). The inverse problem is system of only two equations:

$$\int_0^1 w(x,a,b)dx = 1$$
$$\int_0^1 xw(x,a,b)dx = \frac{\overline{\varepsilon}}{\varepsilon_m} \tag{8}$$

Substituting equations (6) and (7) in equations (8), we have follow equations:

$$\frac{a}{b}(e^b - 1) = 1$$
$$\frac{a}{b}[e^b - \frac{1}{b}(e^b - 1)] = \frac{h\nu}{\varepsilon_m(e^{\frac{h\nu}{kT}} - 1)} \tag{9}$$

Or:

$$a = \frac{b}{(e^b - 1)} \tag{10}$$

$$\frac{1}{(e^b - 1)}[e^b - \frac{1}{b}(e^b - 1)] = \frac{h\nu}{\varepsilon_m(e^{\frac{h\nu}{kT}} - 1)} \tag{11}$$

We can see from equation (10), that, as it must be, we have $a > 0$, at any sign and at any value of the coefficient *b*. For the coefficient *b* we have nonlinear equation (11). The left side of the equation (11), which is marked as *f(b)*, is shown in Fig.1.

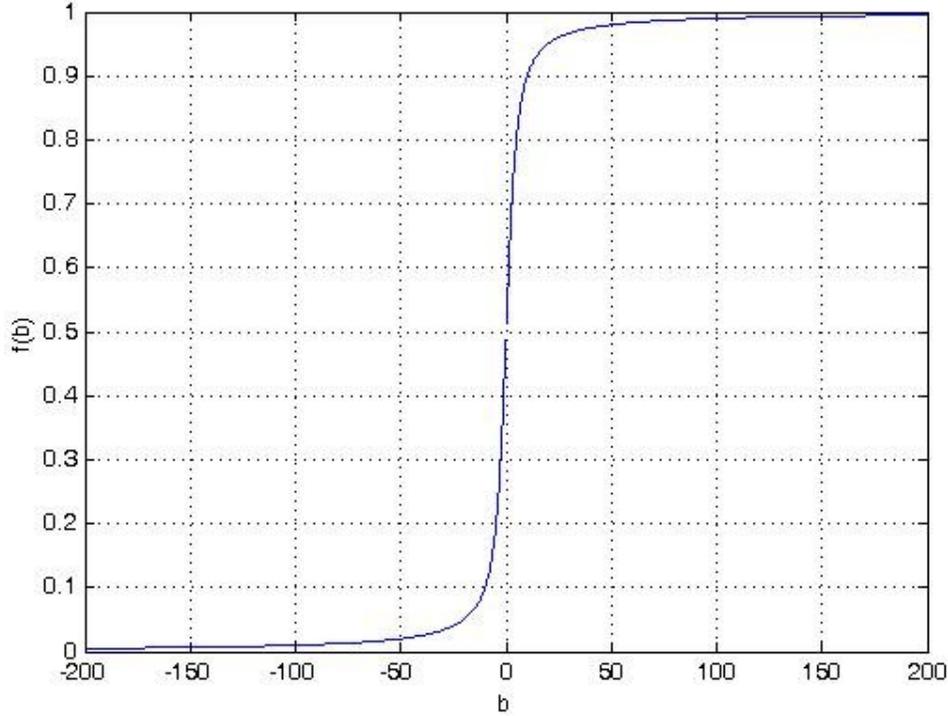

Fig.1. Left side of the equation (11) is function of the coefficient b.

We can see from Fig.1, that the solution $b$ of the equation (11) always exists, because right side of the equation (11) also is in the range [0, 1], i.e. $0 \leq \dfrac{\bar{\varepsilon}}{\varepsilon_m} \leq 1$ . The solution is unique solution.

We can see, that the sign of the coefficient $b$ is negative, if $0 \leq \dfrac{\bar{\varepsilon}}{\varepsilon_m} \leq 0.5$ ('Boltzmann' energy distribution). The sign of the coefficient $b$ is positive, if $0.5 \leq \dfrac{\bar{\varepsilon}}{\varepsilon_m} \leq 1$ ('Anti – Boltzmann' energy distribution).

Let us consider now a hypothesis about maximal wave energy $\varepsilon_m$. We must satisfy the condition:

$$\dfrac{\bar{\varepsilon}}{\varepsilon_m} \leq 1 \qquad (12)$$

It follows from the equation (6), that, at given temperature $T$, maximal average wave energy is $\bar{\varepsilon} = kT$, if $\nu = 0$. On the other side, if $\nu \to \infty$, then $\bar{\varepsilon} \to 0$. Therefore, must be: $\varepsilon_m > kT$.

To satisfy the equation (12) and to explain the photoelectric effect law (see below), we can postulate follow maximal energy for the wave at the frequency $\nu$:

$$\varepsilon_m = kT + h\nu \qquad (13)$$

In the case $kT \ll h\nu$ we can write:

$$\varepsilon_m \cong h\nu \tag{14}$$

Formula (14) permits us give new explanation for the experimental photoelectric effect law. I emphasize here, that the experimental law is not Einstein law:

$$E_k = h\nu - A \tag{15}$$

Where: $E_k$ is kinetic energy of a photoelectron; $A$ is the electron liberation work.
Experimental photoelectric effect law is:

$$E_m = h\nu - A \tag{16}$$

Here $E_m$ is the <u>maximal</u> kinetic energy of a photoelectron.
 We can explain the <u>maximal</u> kinetic energy of a photoelectron in (16) by the <u>maximal</u> energy of wave (14), which is falling on the metal.
 So, we solved inverse problem for Planck formula in the class of Gibbs ensembles. In principle, there is infinite number of energy distributions, which have the formula as the first moment. The choice of only distribution must be governed by additional physical conditions. Therefore, Planck formula can't be used as the basis for the usual simplest theoretical postulate: every electromagnetic wave at the frequency $\nu$ has energy $h\nu$.

## 3. Absence of Bohr – van Leeuwen paradox in classical statistical ensembles of moving charges, in finite phase volume.

As it is noted above, one from the aims of the paper is the proposal to restrict the phase volume for all real physical ensembles. The importance of the proposal I would like to demonstrate by elimination of the Bohr – van Leeuwen paradox in ensembles with finite phase volume.
 Bohr – van Leeuwen theorem consists in follow paradox: classical statistical ensembles of moving charges in external static magnetic field can't have the induced orbital magnetic moment. I.e., the really observed diamagnetism is not possible process [7].
 Let us consider Bohr – van Leeuwen theorem, follow, for example, by [8]. The resulting orbital magnetic moment of moving charges is:

$$\mathbf{p}_{mag} = \frac{1}{2c} \sum_i q_i [\mathbf{r}_i, \mathbf{v}_i] \tag{17}$$

here: $c$ is vacuum light velocity; $q$ is the value of a charge; $\mathbf{r}$ is the coordinate vector of the charge; $\mathbf{v}$ is the velocity vector of the charge.
 The magnetic moment of the ensemble we can to formulate as statistical average value in thermal equilibrium ensemble of point charges in external static magnetic field, without own spin magnetic moments.
 As it is predicted by Maxwell – Boltzmann statistic, the average value of any function is:

$$\overline{f} = \frac{\int_{-\infty}^{\infty} f(\mathbf{r},\mathbf{p}) \exp(-\frac{H(\mathbf{r},\mathbf{p})}{kT}) d\mathbf{r} d\mathbf{p}}{Z} \tag{18}$$

$$Z = \int_{-\infty}^{\infty} \exp(-\frac{H(\mathbf{r},\mathbf{p})}{kT}) d\mathbf{r} d\mathbf{p} \tag{19}$$

Here: **p** is mechanical moment of a charge; $k$ is Boltzmann constant; $T$ is absolute temperature of the ensemble; $\mathcal{H}$ is Hamiltonian of the ensemble.

Emphasize here the infinite limits of the integrals in phase space. Namely the infiniteness will gives us the paradox below.

Let the charges have a weak interactions one with other. Then the charges are statistical independent one from other, and we can to consider only one charge. We will consider the motion of a charge in the plane $Z=0$, with external magnetic field along $Z$ – axis.

Then Hamiltonian of a charge is:

$$\mathcal{H} = \frac{1}{2m}\left[\left(p_x - \frac{q}{c}A_x\right)^2 + \left(p_y - \frac{q}{c}A_y\right)^2\right] \tag{20}$$

Here: **A** is magnetic vector potential; $m$ is a mass of a charge.
We have further:

$$\mathbf{B} = rot\mathbf{A} = (0,0,B) \Rightarrow \mathbf{A} = \frac{B}{2}(-y,x,0) \tag{21}$$

Therefore:

$$\mathcal{H} = \frac{1}{2m}\left[\left(p_x + \frac{q}{2c}By\right)^2 + \left(p_y - \frac{q}{2c}Bx\right)^2\right] \tag{22}$$

Orbital magnetic moment (17) is:

$$\mathbf{P}_{mag} = (0,0,\frac{q}{2c}(xV_y - yV_x)) \tag{23}$$

Velocity vector can be found from the Hamilton mechanics equations:

$$\mathbf{V} = \frac{\partial \mathcal{H}}{\partial \mathbf{p}} \Rightarrow V_x = \frac{1}{m}\left(p_x + \frac{q}{2c}By\right); V_y = \frac{1}{m}\left(p_y - \frac{q}{2c}Bx\right); V_z = 0; \tag{24}$$

Then, we have:

$$\begin{aligned}\mathbf{P}_{mag} &= (0,0,\frac{q}{2c}(x\frac{1}{m}\left(p_y - \frac{q}{2c}Bx\right) - y\frac{1}{m}\left(p_x + \frac{q}{2c}By\right))) = \\ &= (0,0,\frac{q}{2mc}(x\left(p_y - \frac{q}{2c}Bx\right) - y\left(p_x + \frac{q}{2c}By\right)))\end{aligned} \tag{25}$$

We can see from (25), that the component of the magnetic vector along $Z$ – axis is:

$$p_{mag\ z} = -\frac{\partial \mathcal{H}}{\partial B} \tag{26}$$

Or, in the vector view, we have:

$$\mathbf{p}_{mag} = -\frac{\partial H}{\partial \mathbf{B}} \tag{27}$$

Now, the average value of the orbital magnetic moment can be written as:

$$\overline{p_{mag\,z}} = -\frac{\int_{-\infty}^{\infty} \frac{\partial H}{\partial B} \exp(-\frac{H(\mathbf{r},\mathbf{p})}{kT}) d\mathbf{r} d\mathbf{p}}{Z} = kT \frac{\partial \ln Z}{\partial B} \tag{28}$$

Bohr – van Leeuwen theorem tell us:

$$\overline{p_{mag\,z}} = kT \frac{\partial \ln Z}{\partial B} = 0 \tag{29}$$

To proof (29), we must only to change the independent variables in the integral Z :

$$g_x = p_x - \frac{q}{c} A_x;\ g_x = p_y - \frac{q}{c} A_y; \tag{30}$$

It is light to see, that after the substitution, we eliminate the vector potential **A** (or induction flux **B**) from the integral Z . Therefore, the derivative in (29) will be zero. I.e., the Bohr – van Leeuwen theorem is proofed. Diamagnetism is absent.

The paradox is the consequence only of the infinite limits of integral Z, because magnetic vector potential **A** must be included in the integral limits after the substitution (30), if the limits are finite values, i.e. in the case of finite phase volume of the ensemble. Let us consider below the last case.

Let the upper modulus of every coordinate is *L*, and the upper modulus of every component of the mechanical moment is *P*. Then we have:

$$Z = \int_{-L}^{L} dx dy \left[ \int_{-P}^{P} dp_x dp_y \exp\left( -\frac{1}{2mkT} \left[ \left( p_x + \frac{q}{2c} By \right)^2 + \left( p_y - \frac{q}{2c} Bx \right)^2 \right] \right) \right] \tag{31}$$

After the substitution (30), we have:

$$Z = \int_{-L}^{L} dx dy \left[ \int_{-P-\frac{q}{2c}Bx}^{P-\frac{q}{2c}Bx} \exp\left( -\frac{g_y^2}{2mkT} \right) dg_y \int_{-P+\frac{q}{2c}By}^{P+\frac{q}{2c}By} \exp\left( -\frac{g_x^2}{2mkT} \right) dg_x \right] \tag{32}$$

Now, using the Leibniz rule, we have:

$$\frac{\partial Z}{\partial B} = \int_{-L}^{L} dxdy \frac{\partial}{\partial B} \int_{P1P3}^{P2P4} F(g_x, g_y) dg_x dg_y = \int_{-L}^{L} dxdy \begin{bmatrix} \int_{P1P3}^{P2P4} \frac{\partial F(g_x, g_y)}{\partial B} dg_x dg_y + \frac{\partial P_2}{\partial B} \int_{P3}^{P4} F(P_2, g_y) dg_y - \\ -\frac{\partial P_1}{\partial B} \int_{P3}^{P4} F(P_1, g_y) dg_y + \frac{\partial P_4}{\partial B} \int_{P1}^{P2} F(g_x, P_4) dg_x - \\ -\frac{\partial P_3}{\partial B} \int_{P1}^{P2} F(g_x, P_3) dg_x \end{bmatrix}$$

In our case: $\frac{\partial F(g_x, g_y)}{\partial B} = 0$. Then:

$$\frac{\partial Z}{\partial B} = \frac{q}{2c} \int_{-L}^{L} dxdy \begin{bmatrix} y \left[ \exp\left(-\frac{(P + \frac{q}{2c}By)^2}{2mkT}\right) - \exp\left(-\frac{(-P + \frac{q}{2c}By)^2}{2mkT}\right) \right] \int_{-P-\frac{q}{2c}Bx}^{P-\frac{q}{2c}Bx} \exp\left(-\frac{g_y^2}{2mkT}\right) dg_y + \\ + x \left[ \exp\left(-\frac{(-P - \frac{q}{2c}Bx)^2}{2mkT}\right) - \exp\left(-\frac{(P - \frac{q}{2c}Bx)^2}{2mkT}\right) \right] \int_{-P+\frac{q}{2c}By}^{P+\frac{q}{2c}By} \exp\left(-\frac{g_x^2}{2mkT}\right) dg_x \end{bmatrix}$$

(33)

Let the parameters (L, P, B) have appropriate values to satisfy the condition: $g_x \sim g_y \sim 0$. Then, we can to do follow approximation: $\exp(-\frac{g^2}{2mkT}) \cong 1 - \frac{g^2}{2mkT}$.

And we have the final result from (32, 33), by using the linear approximation of the integrals:

$$\frac{\partial Z}{\partial B} = \frac{q}{2c} \int_{-L}^{L} dxdy \left[ y \left[ -\frac{qByP}{mckT} \right] \int_{-P-\frac{q}{2c}Bx}^{P-\frac{q}{2c}Bx} \exp\left(-\frac{g_y^2}{2mkT}\right) dg_y + x \left[ -\frac{qBxP}{mckT} \right] \int_{-P+\frac{q}{2c}By}^{P+\frac{q}{2c}By} \exp\left(-\frac{g_x^2}{2mkT}\right) dg_x \right] =$$

$$= \frac{q}{2c} \int_{-L}^{L} dxdy \left[ y \left[ -\frac{qByP}{mckT} \right] 2P + x \left[ -\frac{qBxP}{mckT} \right] 2P \right] = -\frac{1}{mkT} \left(\frac{q}{c}\right)^2 P^2 B \int_{-L}^{L} dxdy [y^2 + x^2] =$$

$$= -\frac{8}{3} L^4 \frac{q^2}{mc^2 kT} P^2 B$$

(34)

By the same way, from (32) follows the result:

$$Z = 16(LP)^2 \tag{35}$$

Now, from (28) follows average value of the orbital magnetic moment of a charge, in external magnetic field:

$$p_{mag\,z} = -\frac{1}{6}\frac{(qL)^2}{mc^2}B \qquad (36)$$

It is well known formula for the really observed diamagnetism [7].
   Therefore, the Bohr – van Leeuwen theorem is valid only for the statistical ensembles with infinite phase volume, not for real ensembles with finite phase volumes and finite energies.

References:
.
1. *Y.P.Terlezky* Statistical physics. M.: Viscshaja schcola, 1994, 350 pp.
2. *I.P.Basarov* Thermodinamics. M.: Viscshaja schcola, 1991, 376 pp.
3. *A.P.Mosk* //Atomic gases at negative kinetic temperature//Phys.Rev.Lett., **95**, 040403 (2005).
4. *B.H.Lavenda*//Do 'negative' temperatures exist?//J.Phys.A:Math.Gen. **32** (1999) 4279-4297.
5. *N.F.Ramsey*//Thermodynamics and statistical mechanics at negative absolute temperatures//Phys.Rev. **103**, 1, 20 – 28 (1956).
6. *N.I.Kalitievskij* Wave optick. M.: Viscshaja schcola, 1978, 349 pp.
7. *S. V Vonsovskii* Magnetism. M.: Nauka, 1971, p.70.
8. *R. Becker* Teoria electrichestva, v.2, Electronnaja teoria. M.: GITTL, 1941, 389 pp.